\documentclass[11pt,reqno]{amsart}

\usepackage[parfill]{parskip}  
\usepackage[english]{babel}

\usepackage{amsmath}
\usepackage{amsfonts}
\usepackage{amssymb}
\usepackage{amsthm}
\usepackage[pdftex]{graphicx}
\usepackage{epstopdf}
\usepackage{enumitem}
\usepackage{geometry}
\usepackage{hyperref}
\usepackage{tikz}
\usepackage[all]{xy}
\usepackage{multirow}
\usepackage{anysize}
\usepackage{color}
\usepackage{cite}
\usepackage{lipsum}
\usepackage{lineno}
\usepackage{setspace}

\newcommand\blfootnote[1]{%
  \begingroup
  \renewcommand\thefootnote{}\footnote{#1}%
  \addtocounter{footnote}{-1}%
  \endgroup
}

\newcommand{\beq}{\begin{equation}}
\newcommand{\eeq}{\end{equation}}

\marginsize{3.2cm}{3.2cm}{3.2cm}{3.2cm}

\title{Benevolent characteristics promote cooperative behaviour among humans}

\begin{document}

\maketitle

\begin{center}VALERIO CAPRARO*\blfootnote{* These authors contributed equally to this work.}\footnote{Corresponding Author. Valerio Capraro, email: V.Capraro@soton.ac.uk, Telephone Number: +44 747 3055508}\\Department of Mathematics\\University of Southampton\\Building 54\\ University Road\\ SO17 1BJ\\ United Kingdom\\

\end{center}
\bigskip
\begin{center}CONOR SMYTH*\\Department of Mathematics\\University of Southampton\\Building 54\\ University Road\\ SO17 1BJ\\ United Kingdom\\

\end{center}
\bigskip
\begin{center}KALLIOPI MYLONA\\Department of Mathematics\\University of Southampton\\Building 54\\ University Road\\ SO17 1BJ\\ United Kingdom\\

\end{center}
\bigskip
\begin{center}GRAHAM A. NIBLO\\Department of Mathematics\\University of Southampton\\Building 54\\ University Road\\ SO17 1BJ\\ United Kingdom\\

\end{center}

Keywords: cooperation, prosocial behaviour, benevolence, altruism, inequity aversion.

\doublespacing

\begin{abstract}
Cooperation is fundamental to the evolution of human society. We regularly observe cooperative behaviour in everyday life and in controlled experiments with anonymous people, even though standard economic models predict that they should deviate from the collective interest and act so as to maximise their own individual payoff. However, there is typically heterogeneity across subjects: some may cooperate, while others may not. Since individual factors promoting cooperation could be used by institutions to indirectly prime cooperation, this heterogeneity raises the important question of who these cooperators are. We have conducted a series of experiments to study whether benevolence, defined as a unilateral act of paying a cost to increase the welfare of someone else beyond one's own, is related to cooperation in a subsequent one-shot anonymous Prisoner's dilemma. Contrary to the predictions of the widely used inequity aversion models, we find that benevolence does exist and a large majority of people behave this way. We also find benevolence to be correlated with cooperative behaviour. Finally, we show a causal link between benevolence and cooperation: priming people to think positively about benevolent behaviour makes them significantly more cooperative than priming them to think malevolently. Thus benevolent people exist and cooperate more.
\end{abstract}

\section*{Introduction}

Two or more people cooperate if they pay an individual cost in order to increase the welfare of the group. The canonical economic model, assuming people care only about their own welfare, predicts that they should not cooperate: the incentive to minimise individual cost causes people to act selfishly. In reality the opposite behaviour is often observed. In personal relationships, workplace collaborations, political participation, and concerning global issues such as climate change, examples of cooperation are manifold, and fostering cooperation has been show to have a number of important applications\cite{C09, ZM, Tr, Ax-Ha, MSK, DH05, LK06, No06, Ra-No, Ca}. 

Classical studies have been focussed on punishing of defectors\cite{BR, FG1, FG2, GIR}, increasing the reputation of cooperators \cite{MSK, PB, MSKM}, and the interplay between these two mechanisms\cite{AHV,RM,SSW,HS}. While these approaches have been successfully shown to enforce cooperation, and punishment has been adopted by most countries to sanction defectors, their weakness is their cost to not only the punisher and the punished, but to the third party tasked with rewarding those with increased reputation. The principle is: if we want to increase cooperation, someone must pay a cost. 

In this light, it becomes important to find less expensive ways to sustain cooperation and it is here that individual factors may play a crucial role. Assume individual factor $X$ is known to promote cooperation, then creating an environment which favours factor $X$ will also promote cooperation. Existence of one or more such factors is suggested by the numerous experimental studies showing that humans do tend to behave cooperatively, even in anonymous, isolated environments where communications or long-term strategies are not allowed\cite{C,GH,Ze,CJR}. These studies have shown that humans are heterogeneous: some may cooperate, while others may not. If so, who are the cooperators?

A growing body of literature is trying to provide answers to this question, by investigating what factors promote cooperation in one-shot social dilemma games, such as the Public Goods game and the Prisoner's dilemma. In the Public Goods game, $N$ agents are endowed with $y$ monetary units and have to decide how much, if any, to contribute to a public pool. The total amount in the pot is multiplied by a constant and evenly distributed among all players. So, player $i$'s payoff is $y-x_i+\alpha(x_1+\ldots+x_N)$, where $x_i$ denotes $i$'s contribution and the `marginal return' $\alpha$ is assumed to belong to the open interval $(\frac1N,1)$. In the Prisoner's dilemma, two agents can either cooperate or defect. To cooperate means paying a cost to give a greater benefit to the other player; to defect means doing nothing.

Previous experimental studies have investigated the role of intuition and altruism on cooperation (see Box 1 for definitions) in one-shot anonymous Public Goods games and Prisoner's dilemma games. Intuitive actions are induced by either exerting time pressure on subjects or priming them towards intuition versus deliberate reflection\cite{RGN,R,PR,RKT}. While it is generally accepted that intuition favours cooperation through the Social Heuristics Hypothesis\cite{RGN}, the correlation between altruism and cooperation is still unclear: one study did not find any correlation between altruism and cooperation in a subsequent one-shot Public Goods game\cite{BEN}, while another study found a positive correlation between altruism and cooperation in a precedent one-shot Prisoner's dilemma\cite{CJR}.

Altruism is formally defined as unilaterally paying a cost $c\geq0$ to give a benefit $b$ to another and is traditionally measured using a Dictator game\cite{CJR, PR,DFR,HK,BEN}. Here a dictator is given an endowment $x>0$ and must then decide how much, if any, to donate to a recipient who was given nothing. The recipient has no input in the process and simply accepts the donation. Givings in the Dictator game are usually considered as an appropriate measure of altruism\cite{BG06,BG07,CG08,E} and recent experiments have shown that indeed they positively correlate to altruistic acts in real-life situations\cite{FP}. 

Experiments on the Dictator game typically present a bimodal distribution. Participants tend to either act selfishly or act so as to decrease inequity between players. Consider the scenario where Player 1 is given \$10 dollars and must then decide how much if any to donate to a second anonymous player. In most cases Player 1 decides to selfishly keep all of the money, or to donate half to Player 2, and so reduce the inequity between the two players. There is a third scenario that occurs, although rarely. Here Player 1 decides to donate more to Player 2 than to keep for herself. In some cases players have been known to donate the entire sum. The act of increasing the other payoff beyond your own will be called `benevolence'. It is likely that this behaviour is not observed more often in the Dictator game as its design effectively penalises altruism. If cost were less of a factor perhaps benevolence would be more prevalent.

\colorbox{yellow}{
  \parbox{\textwidth}{
   \textbf{Cooperation.} Two or more people cooperate if they pay an individual cost to give a greater benefit to the group.\\
   \textbf{Altruism.} A person acts altruistically if he \emph{unilaterally} pays a cost $c\geq0$ to increase the benefit of someone else. More formally, Player 1 is altruist towards Player 2 if he prefers the allocation $(x_1-c,x_2+b)$ to the allocation $(x_1,x_2)$, where $c\geq0$ and $b>0$.\\
   \textbf{Benevolence.} A person acts benevolently if he \emph{unilaterally} pays a cost $c\geq0$ to increase the benefit of someone else beyond one's own. More formally, Player 1 is benevolent towards Player 2 if he prefers the allocation $(x_1-c,x_2+b)$ to the allocation $(x_1,x_2)$, where $c\geq0$, $b>0$, and $x_2+b>x_1-c$.\\
   In sum, the main difference between cooperation and altruism is that altruism is unilateral: there is no way to get rewarded. Another difference is that we allow altruist action at negligible cost. In other words, the important part is to create a benefit to someone else without getting anything back. Benevolence is an extreme form of altruism, where the final result of the act is that the recipient has a larger payoff of the actor.
} }

Examples of benevolence in everyday life abound. The sharing of one's food causing the sharer to go hungry, campaigning on behalf of a VIP in order to promote their agenda, or something as trivial as `liking' or sharing a status on social networks so as to increase the reputation of another.

In this paper, we have designed a game that allows players to choose actions that are malevolent, inequity averse or benevolent, all at minimal cost. More specifically, we give an endowment $L>0$ to Player 1 that she keeps regardless of any subsequent choice. She has to then decide how much, between $0$ and $H\geq L$ to donate to Player 2. To donate $0$ will be referred as a malevolent act; to donate $L$ will be referred as inequity aversion; to donate more than $L$ will be referred as benevolence.

This form of benevolence, though costless, increases the inequity among people and so it is predicted not to exist by the widely used inequity aversion models\cite{FS,BO}. Thus, as a first step of our program, we have conducted an experiment, using this new economic game, to show that most people act in a benevolent way even when it is made clear that there is no possibility of an indirect reward. We next move to investigate our main research question: Is benevolence one of those individual factors favouring cooperative behaviour? With this is mind, as a second step, we have asked whether benevolence is correlated to cooperative behaviour. We have found that benevolence positively correlates with cooperation in a number of different settings, and with different payoffs. Finally, in our third study, we have showed the causal link between benevolence and cooperation: priming people towards benevolence versus malevolence results in a significant increase of cooperative behaviour. 

These results allow us to conclude that benevolence is an individual factor possessed by many people and that it is among those factors promoting cooperative behaviour. Although this observation contradicts inequity aversion models, other theories could be used to explain it. For instance, the tendency to maximise the total welfare and adherence to social norms can explain the existence of benevolence and its correlation with cooperative behaviour. We refer the reader to the Discussion section for more details.

\section*{Study 1. benevolence exists.}

Inequity aversion models\cite{FS,BO} are based on the assumption that humans have a tendency to mitigate payoff differences. Since benevolence, measured using the game described below, increases payoff difference between the actor and the recipient, these models predict that it does not exist. Thus, as a first step of our program, we make us sure that benevolence does actually exist. Moreover, we test whether people trust in the benevolence of others and, to this end, we have introduced a second player who has to gamble on the first player's level of benevolence. Among the several different ways one can formalise this strategic situation through an economic game, we have adopted a particularly simple one, formally described in Box 2.

\colorbox{yellow}{
  \parbox{\textwidth}{
   \textbf{BT$(L,H)$.} Player 1 is given an amount $L\geq0$ of dollars which she keeps regardless her choice. She then must choose an amount of dollars between $0$ and $H\geq L$. Player 2 has to decide a number between 0 and $H$, as well. If Player 2's choice, say $t$ (as in trust), is smaller than or equal to Player 1's choice, say $b$ (as in benefit or benevolence), then Player 2 gets $t$ dollars, otherwise he gets nothing. So player 1's decision corresponds to the maximum amount of dollars she allows player 2 to make, while player 2's choice is a measure of his trust in Player 1's benevolence.}
}

We recruited US subjects to play BT$(\$0.10,\$1)$ using the online labour market Amazon Mechanical Turk\cite{PCI,HRZ,R12,TSW}. After explaining the rules, we asked a series of comprehension questions to make sure they understood the game. These questions were formulated to make very clear the duality between harming and favouring the other player at zero cost for themselves. Players failing any of the comprehension questions were automatically screened out. We refer the reader to the Supporting Information for full experimental instructions.

\begin{figure} 
   \centering
   \includegraphics[scale=0.50]{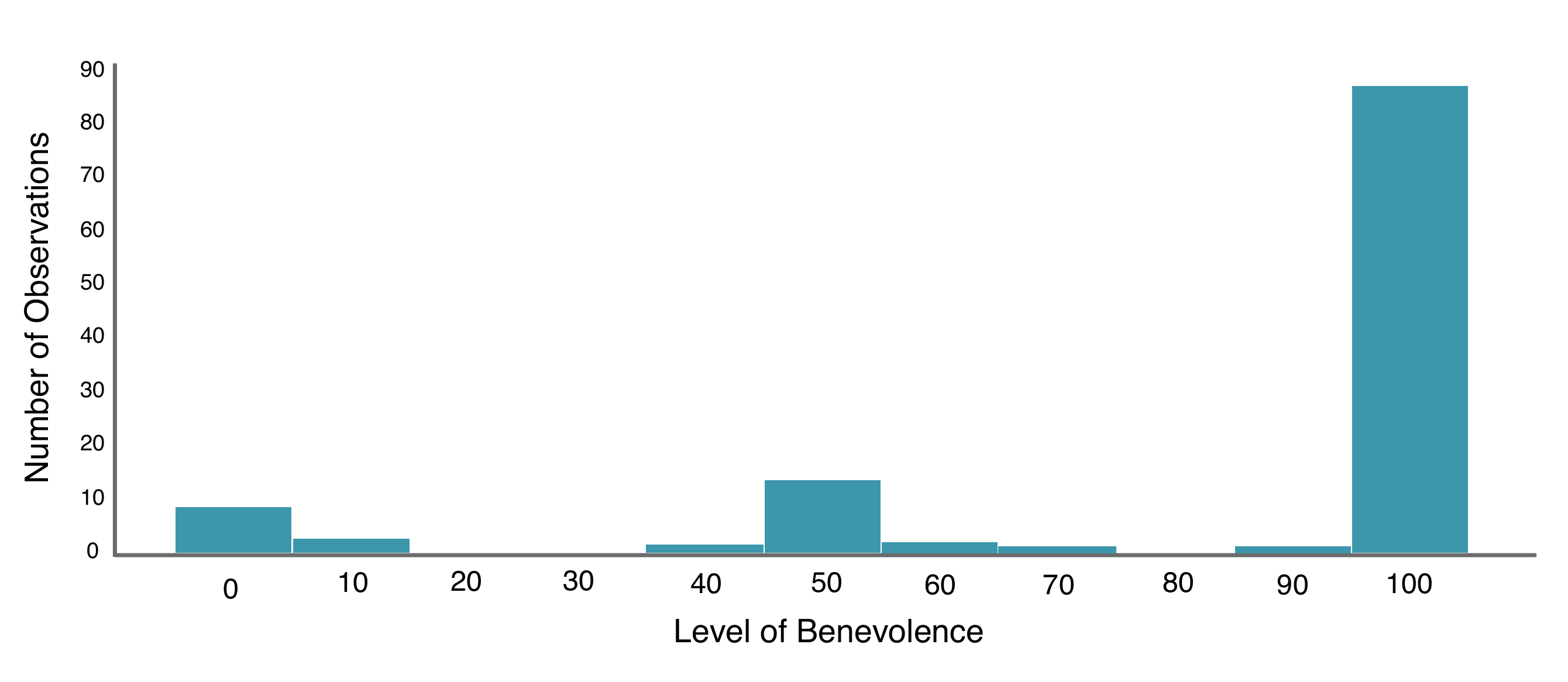} 
   \caption{Distribution of choices in BT$(\$0.10,\$1)$ of those people acting as Player 1. Only 12 out of 123 participants acted in a malevolent or inequity averse way; all others acted benevolently with a large majority of participants acting in a perfectly benevolent way.}
   \label{fig:goodness}
\end{figure}

A total of 247 subjects passed all comprehension questions. Among the 123 subjects who played as Player 1, we find that only 12 participants chose a strategy $b\leq\$0.10$ (9 malevolent and 3 inequity averse people). All others chose $b\geq\$0.40$ and about $60\%$ of the subjects acted in a perfectly benevolent way, choosing $b=\$1$ and so maximising the inequity between themselves and the others (see Figure \ref{fig:goodness}).

By looking at the 124 subjects who played as Player 2, we find that subjects tended to trust in the benevolence of others, although we find a general tendency to underestimate it: while the average `benevolence' was $\$0.82$, the average `trust' was only $\$0.63$. The Mann-Whitney test confirms that these samples most likely come from different distributions ($P<.0001$). We have also conducted a similar experiment with BT$(\$0.10,\$0.10)$ with 133 US subjects acting as Player 1 and 142 as Player 2. By comparing the results in BT$(\$0.10,\$1)$ with those in BT$(\$0.10,\$0.10)$ we find that, after the obvious rescaling, benevolence and trust do not seem to depend on the maximum payout $H$. (Mann-Whitney test: $P=0.538$ in case of benevolence; $P=0.981$ in case of trust). 
\section*{Study 2. benevolence is positively correlated with cooperation.}

To study correlation between cooperation and benevolence, and cooperation and trust, we designed a battery of four two-stage games. Participants first played a BT game and then a standard Prisoner's dilemma PD$(b,c)$ with cost $c=\$0.10$ and benefit $b=\$0.25$. In our PD, two players must choose to either either cooperate or defect: to defect means keeping $c$, while to cooperate means giving $b$ to the other player. The strategic situation faced by the participants is summarised in Box 3.

\colorbox{yellow}{
  \parbox{\textwidth}{
   \textbf{T1.} Subjects first play GT$(\$0.10,\$0.10)$ as Player 1 and then play PD$(\$0.25,\$0.10)$.\\
   \textbf{T2.} Subjects first play GT$(\$0.10,\$1)$ as Player 1 and then play PD$(\$0.25,\$0.10)$.\\
   \textbf{T3.} Subjects first play GT$(\$0.10,\$0.10)$ as Player 2 and then play PD$(\$0.25,\$0.10)$.\\
   \textbf{T4.} Subjects first play GT$(\$0.10,\$1)$ as Player 2 and then play PD$(\$0.25,\$0.10)$.
  }
}

Again we recruited US subjects using AMT and asked qualitative comprehension questions to make sure they understood the game. 

\begin{figure} 
   \centering
   \includegraphics[scale=0.50]{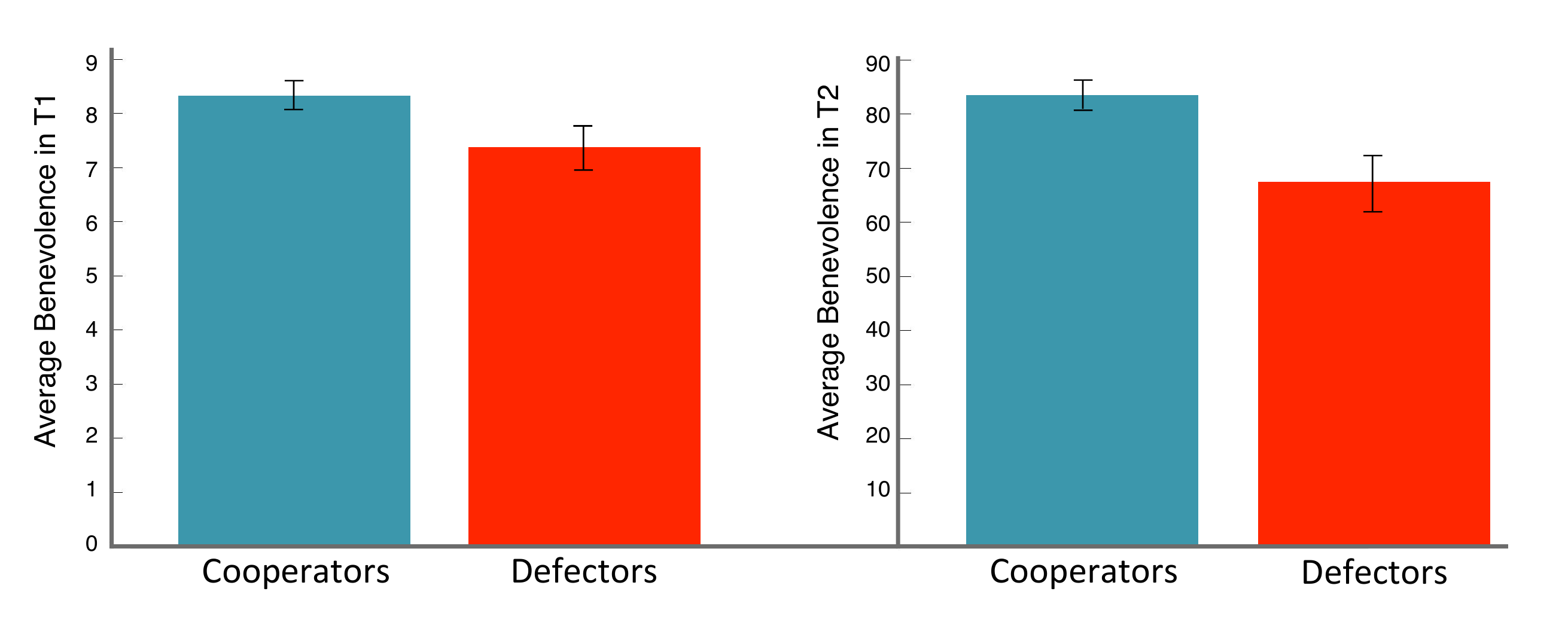} 
   \caption{Average level of benevolence of cooperators and defectors in T1 and T2, with error bars representing the standard error of the mean. In both treatments benevolence is positively correlated with cooperation.}
   \label{fig:goodness-coop}
\end{figure}

A total of 385 subjects, nearly evenly distributed among the four treatments, passed all comprehension questions. Figure \ref{fig:goodness-coop} shows the average benevolence of cooperators and defectors in T1 and T2. Benevolence seems positively correlated with cooperation in both treatments. To confirm this, we use logistic regression to predict defection or cooperation as the dependent variable. We find that the correlation between benevolence and cooperation is borderline significant in T1 (coeff $=0.175$, $P=0.054$) and significant in T2 (coeff $=0.02$, $P=0.019$). On the other hand, we find that trust affects cooperation only for $H=0.1$ (coeff $=0.167$, $P=0.016$) and does not for $H=1$ (coeff $=-0.001$, $P=0.887$). 

\section*{Study 3. Priming benevolence promotes cooperation.}

In the previous studies we have shown that benevolence exists and is positively correlated to cooperation. Here we show the causal link between benevolence and cooperation. To do this we use an experimental design similar to that used in \cite{RGN} to show the causal link between intuition and cooperation: we prime participants towards benevolence or malevolence before playing a Prisoner's dilemma. Specifically, we have conducted three more treatments, as described in Box 4.

\colorbox{yellow}{
  \parbox{\textwidth}{
   \textbf{T5.} After entering the game, participants see a screen where we define benevolence as giving a benefit to someone else at negligible cost to themselves. Subjects are then asked to write a paragraph describing a time when acting benevolently led them in the right direction and resulted in a positive outcome for them.  Alternatively, they could write a paragraph describing a time when acting malevolently led them in the wrong direction and resulted in a negative outcome for them. After this, they are asked to play PD$(\$0.25\$,0.10)$. \\
   \textbf{T6.} This treatment is very similar to T5, with the only difference that subjects are primed towards malevolence. We first define malevolence as an unkind act towards someone else with no immediate benefit for themselves and then we ask participants to write a paragraph describing a time when acting benevolently led them in the wrong direction and resulted in a negative outcome for them.  Alternatively, they could write a paragraph describing a time when acting malevolently led them in the right direction and resulted in a positive outcome for them. \\
   \textbf{T7.} This is a baseline treatment, where participants enter the game and are immediately asked to play PD$(\$0.25\$,0.10)$, using literally the same instructions as in T5 and T6, in order to avoid framing effects.
   
  }
}

Also for this study, we recruited US subjects using AMT. In order not to destroy the priming effect we decided not to ask for comprehension questions before the Prisoner's dilemma in T5 and T6. Further, we asked no comprehension questions in T7 so as not to bias any baseline measurement. To control for good quality results we used other techniques (see Appendix for full experimental details). In particular, at the end of the experiment we asked the players to describe the reason of their choice. This, together with the descriptions of benevolent or malevolent actions, allowed us to manually exclude from the analysis those subjects who did not take the game seriously or showed a clear misunderstanding of the rules of the game, as sometimes happens in AMT experiments. We excluded from our analysis 11 subjects.

\begin{figure} 
   \centering
   \includegraphics[scale=0.60]{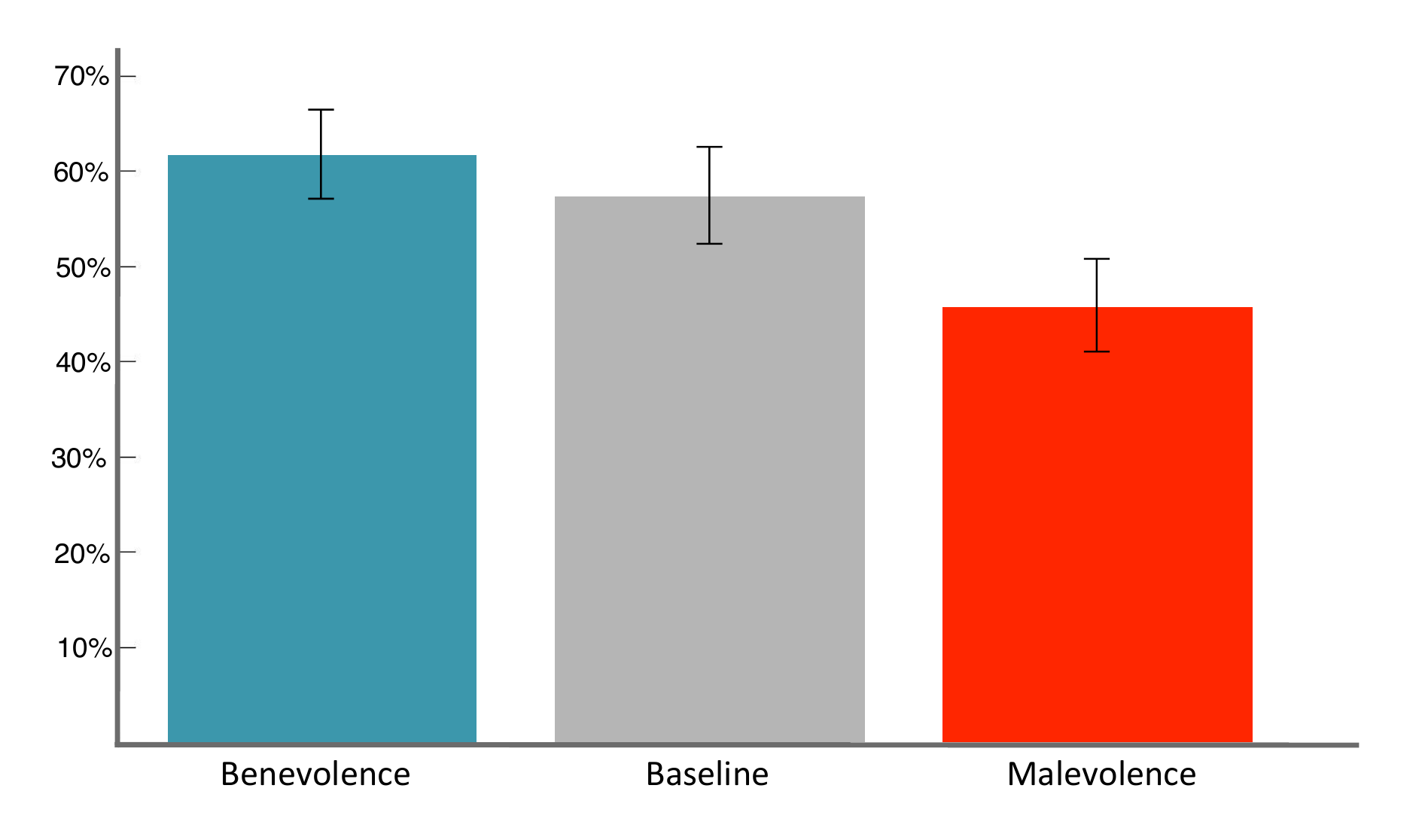} 
   \caption{Average cooperation in each of the three treatments. Participants primed to act benevolently were significantly more likely to cooperate than than those primed to act malevolently. The level of cooperation of unprimed participants lies between those of the primed groups and cannot be statistically separated from either.}
   \label{fig:benMal}
\end{figure}

300 US subjects, nearly evenly distributed among the three treatments, participated to our third study and passed our manual screening. As Figure \ref{fig:benMal} shows, the trend is in the expected direction. $62\%$ of the participants cooperated in T5, far more than that in T6 ($46\%$). Pearson's $\chi^2$ test confirms that the difference is statistically significant ($\chi^2(1)=5.425, P=0.020$). The baseline treatment lies just in between, with a percentage of cooperation of $58\%$. However, the difference is not statistically significant with neither of the other two treatments (T5 vs T7, $\chi^2(1)=0.394, P=0.530$; T6 vs T7, $\chi^2(1)=2.629, P=0.105$). 
\section*{Discussion}
Benevolence, that is paying a (potentially zero) cost to increase someone else's welfare beyond that of your own, is predicted not to exist by the widely used inequity aversion models\cite{FS,BO}, which are indeed founded on the idea that humans have a tendency to moderate payoff differences. Contrary to this prediction, our Study 1 shows that most people act in a benevolent way, at least when the cost of the action is zero, and that most people trust in the benevolence of others. Existence of benevolence might be seen as surprising also in light of other experimental results showing that people are often willing to pay a cost to decrease the benefit of a richer partner \cite{DF}. The explanation of these apparently contradictory results most likely relies in social norms: most people think that to be benevolent and to restore equity when a situation of inequity is artificially presented as \emph{status quo} without any reason as in \cite{DF} are both the `right things to do'.

Looking for individuals factors promoting cooperative behaviour, we have then asked whether benevolence is positively correlated with cooperative behaviour. Our Study 2 shows that: benevolent people not only exist, but they are more likely to cooperate in a subsequent Prisoner's dilemma. This provides evidence that benevolence is one of those individual factors favouring cooperation. 

Our Study 3 strengthens this conclusion by directly showing a causal link between benevolence and cooperation. Priming people to think about benevolence in a positive way or about malevolence in a negative way makes them significantly more cooperative than priming them to think about benevolence in a negative way or about malevolence in a positive way. 
The fact that the level of cooperation can experimentally be manipulated in such a way connects to the important question of whether priming people towards benevolence can be used for instance by companies as a way to increase cooperation among employees or by countries to increase cooperation among citizens, and to what extent.

As we have mentioned, our results are not consistent with inequity aversion models. However, a number of other theories could explain both the existence of benevolence and its positive correlation with cooperative behaviour. Several experimental studies have shown that many people act so as to maximise the total welfare\cite{Ch-Ra, ES} and some of the most recent mathematical models of human behaviour are indeed based on postulating this tendency\cite{Ch-Ra,Ca,CVPJ,BC}. This predisposition might explain why benevolent people exist and are more cooperative: our results might be due to a number of people attempting to maximise the total welfare. Other scholars suggest that social norms shape most of cooperative behaviour\cite{FF,T13}. Though social norms varies across cultures, it is possible that to be benevolent and to be cooperative are both seen as the `right things to do' by part of the US population. From this perspective, our results could be driven by a number of people attempting to act according to the social norm they adhere to. 

We conclude by saying that our results do not imply that defectors are never benevolent. As Figure \ref{fig:goodness-coop} shows, defectors were substantially more benevolent than predicted by inequity aversion models, but significantly worse than cooperators. Benevolence thus seems far more transversal than cooperation and suggests the following question. What evolutionary pressures select for benevolence? Together with the aforementioned theories, several others, such as warm-glow giving\cite{A} and the Social Heuristics Hypothesis\cite{RGN}, offer qualitative explanations. Andreoni's warm-glow giving theory states that (some) humans receive utility from the fact itself of giving; the SHH instead builds on the idea that everyday life interactions are often repeated and a benevolent act today may be rewarded tomorrow. It is then possible that people internalise benevolence in their everyday life and use it as a default strategy in the lab.  

In conclusion, we have found that benevolence exists and it is positively correlated to cooperation. However, the ultimate reason why benevolence exists and why it is correlated with cooperation is far from being clear. It is therefore an important question for further research and is likely to be challenging because it clearly connects to some of the most basic open problems of human social behaviour.

\section*{Acknowledgements}
We thank Elena Simperl for discussions about the experimental design and Christian Hilbe and David G. Rand for helpful comments on early drafts.

\section*{Materials and Methods}

We recruited US subjects using Amazon Mechanical Turk and randomly assigned them to one of seven experiments using economic games. Treatments are described in the Main Text and full instructions are given in the Appendix. In four treatments, participants were informed that comprehension questions would be asked after the instructions of each game and that they would be automatically eliminated if they failed to correctly answer them. Comprehension questions were formulated in such a way to make very clear the duality between harming and favouring the other player (in case of the BT game), and between maximising one's own payoff and maximising the total welfare (in case of the PD). A total of 385 subjects passed all comprehension questions. Participants were also informed that computation and payment of the bonuses would be made at the end of the experiment. So, importantly, in each treatment, participants played the second game without knowing the outcome of the first. The structure of the remaining three treatments was such that we could not ask for comprehension questions. However, to control for good quality result we used other techniques, such as asking participants to describe the strategy used. This allowed us to manually eliminate from the analysis those people who showed a clear misunderstanding of the game rules. No deception was used. These experiments were approved by the Southampton University Ethics Committee on the Use of Human Subjects in Research. For further details of the experimental methods, see appendix.

\section*{Appendix: Experimental setup}

We recruited US subjects using the online labor market Amazon Mechanical Turk (AMT) \cite{PCI, HRZ,R12}. As in classical lab experiments, AMT workers receive a baseline payment and can earn an additional bonus depending on how they perform in the game. AMT experiments are easy to implement and cheap to realise, since AMT workers are paid a substantially smaller amount of money than people participating in physical lab experiments. Nevertheless, it has been shown that data gathered using AMT agree both qualitatively and quantitatively with data collected in physical labs \cite{HRZ,R12,SW,RAC}.

Our experiment was conducted in two sessions. The first session corresponds to Study 1 and Study 2 in the main text, while the second session corresponds to Study 3.

\subsection*{First Session}
Subjects were paid a $\$0.20$ show-up fee for participating and then randomly assigned to one of eight treatments. 

\textbf{T1.} Subjects first play BT$(\$0.10,\$0.10)$ as Player 1 and then play PD$(\$0.25,\$0.10)$.\\
\textbf{T2.} Subjects first play BT$(\$0.10,\$1)$ as Player 1 and then play PD$(\$0.25,\$0.10)$.\\
\textbf{T3.} Subjects first play BT$(\$0.10,\$0.10)$ as Player 2 and then play PD$(\$0.25,\$0.10)$.\\
\textbf{T4.} Subjects first play BT$(\$0.10,\$1)$ as Player 2 and then play PD$(\$0.25,\$0.10)$.\\

Each treatment consists of two economic games. Subjects were informed that computation and payment of the bonuses would be made at the end of the experiment. So, in each treatment, participants played the second game without knowing the outcome of the first. The reason for doing this is that we want to minimise spill over effects due to the fact that subjects in T1-T4 may behave more or less cooperatively in the second game depending on whether or not they have been recipient of a benevolent act in the first game.

A major issue with AMT is that workers try to play multiple times in order to get a larger bonus and/or play randomly in order to complete the task as soon as they can. To control for these issues is very easy using the Survey builder Qualtrics. At the very beginning of the task we asked for the Turk ID and we automatically excluded from the task people who had already completed it. To address the second issue, we asked comprehension questions for each of the economic games and we automatically excluded from the task workers who fail to correctly answer them.  We stress the fact that we decided not to ask technical questions, such as computing the expected payoff of a strategy profile, and preferred to ask qualitative questions in order to make clear the dualities between making a benevolent action or not (in the BT game) or between maximising the total welfare versus maximising one's own payoff (in the PD).

385 US subjects, nearly evenly distributed among the eight treatments, answered correctly all comprehension questions. 

Here we report the full instructions for Treatment 1. The instructions of the other treatments were absolutely identical, a part from obvious changes in the parameters. The first two screens do not contain any information about the game and serve us only as control to avoid multiple plays from the same subject and lazy participants who can increase randomness on our data.

\textbf{Screen 1.} In the first screen, participants were welcomed to the game and asked to type their worker ID. This allows us to automatically exclude workers who have already completed the task.

\textbf{Screen 2.} In the second screen, we asked the participant to transcribe a relatively long neutral piece of text. This allows us as to tell computers and humans apart (CAPTCHA) and, at the same time, to exclude lazy workers and minimise randomness in our data. We used a meaningless neutral text in order to avoid framing effects.

In the third screen, people entered the real game. Here is the exact instructions we used.

\textbf{Screen 3.} Welcome to this HIT. This HIT has two parts. We will tell you about the second part after you have completed the first one. For your participation in this HIT, you receive $\$0.20$. You also can earn up to $\$1.20$ as a bonus depending on the group to which you are assigned. Full details will be given in the following pages. The size of your bonus depends on the decisions you will make in the tasks that follow but also on the decisions of another anonymous MTurk participant, with whom you are paired. Your bonus will be paid within 10 days of the completion of the HIT. You can opt out at any time although if you choose to exercise this option before completing the HIT then you will not receive any payment or bonus. You will be told about the outcome of all parts of the HIT at the same time that your bonus is paid. If you answer any comprehension question incorrectly the survey will end and unfortunately you will not receive any payment or bonus. With this in mind, do you wish to continue? 

Here participants could either continue or end the survey, clicking on the corresponding button. Participants who decided to continue were directed to the next screen.

\textbf{Screen 4.} This is the first part of the HIT. You have been paired with another anonymous participant, and you now have a decision to make. Your decision will NOT affect how much money you earn, but it may affect the other participant's income. Along with your payment for participating in this HIT you are given 10 additional cents as a bonus, which you will keep regardless of what follows (congratulations!), the other participant has been given nothing. You have to choose an amount between 0c and 10c corresponding to the maximum amount of money the other participant can make. The other participant will try to guess your choice. If the other participant's guess is smaller than or equal to the number you have chosen, then they will earn an amount of money equal to that guess. Otherwise they will get nothing. The other person is REAL and will really make a decision. Now we will ask you some questions to make sure you understand the task. You MUST answer all questions correctly to receive any payment or bonus. If you answer incorrectly the survey will terminate and you will not receive a redemption code. There is no incorrect answer when you are asked to make your actual decision.
\begin{enumerate}
\item Which of the following choices made by YOU favours the OTHER PARTICIPANT the most? 
\item Which of the following choices made by YOU minimises the amount the OTHER PARTICIPANT can win?
\end{enumerate}

In both questions, participants could select all possible integers between 0 and 10, in T1, or all possible multiples of ten between 0 and 100, in T2. Questions contained a Skip Logic, that is a program that automatically ends the survey if the answer is not correct.

\textbf{Screen 5.} Now it's time to really make your decision. Choose an amount from the following list.

The list of possible choices in T1 contained all integers between 0 and 10, while in T2 contained all multiples of 10 between 0 and 100.

\textbf{Screen 6.} This is the second part of the HIT. You are paired with another anonymous participant. The amount of money you earn depends on your choice AND on the choice of the other participant. You and the other participant are both given 10c and you both have the same decision to make. You can either keep the 10 cents or give it all to the other participant who will also receive an extra 5c from us for a total of 15c. So if you both keep the money, you both earn 10c. If you both give all of your money, you will both earn 15c. If you keep all of your money and the other participant gives all of their money, you will earn 25c. If you give all of your money and the other participant keeps all of theirs, then you will earn nothing. The other person is REAL and will really make a decision. Now we will ask you some questions to make sure you understand the task. You MUST answer all questions CORRECTLY to receive any payment or bonus. If you answer incorrectly the survey will terminate and you will not receive a redemption code. There is no incorrect answer when you are asked to make your actual decision.
\begin{enumerate}
\item If the other participant chooses to `Keep', which choice made by YOU maximizes YOUR bonus?
\item If the other participant chooses to `Give', which choice by YOU maximizes the OTHER PARTICIPANT's bonus?
\item If you choose to `Give', which choice by the OTHER PARTICIPANT maximizes the OTHER PARTICIPANT's bonus?
\item If you choose to `Keep', which choice by the OTHER PARTICIPANT maximizes YOUR bonus?
\end{enumerate}
In all questions, participants could select either `keep' or `give'. All questions contained a Skip Logic, that is a program that automatically ends the survey if the answer is not correct.

\textbf{Screen 7.} Now it's time to make your decision. What is your choice?

Here participants were asked to either `keep' or `give'. No intermediate choices as in \cite{CJR} were allowed. Following this screen, we asked demographic questions. A final screen, providing a completion code to claim for their payment, concluded the survey.

\subsection*{Second Session} 
Subjects were paid a $\$0.20$ show-up fee for participating and then randomly assigned to one of three treatments.

   \textbf{T5.} After entering the game, participants see a screen where we define benevolence as giving a benefit to someone else at negligible cost to themselves. Subjects are then asked to write a paragraph describing a time when acting benevolently led them in the right direction and resulted in a positive outcome for them.  Alternatively, they could write a paragraph describing a time when acting malevolently led them in the wrong direction and resulted in a negative outcome for them. After this, they are asked to play PD$(\$0.25\$,0.10)$. 
   
   \textbf{T6.} This treatment is very similar to T5, with the only difference that subjects are primed towards malevolence. We first define malevolence as an unkind act towards someone else with no immediate benefit for themselves and then we ask participants to write a paragraph describing a time when acting benevolently led them in the wrong direction and resulted in a negative outcome for them.  Alternatively, they could write a paragraph describing a time when acting malevolently led them in the right direction and resulted in a positive outcome for them. 
   
   \textbf{T7.} This is a baseline treatment, where participants, after entering the game, are immediately asked to play PD$(\$0.25\$,0.10)$, using literally the same instructions as in T5 and T6, in order to avoid framing effects.
   
In order not to destroy the priming effect we decided not to ask for comprehension questions before the Prisoner's dilemma in T5 and T6. Further, we asked no comprehension questions in T7 so as not to bias any baseline measurement. To control for good quality results we asked the players to describe the reason of their choice. This, together with the descriptions of benevolent or malevolent actions allowed us to manually exclude from the analysis those subjects who did not take the game seriously or showed a clear misunderstanding of the rules of the game. We excluded 11 subjects from the analysis. 300 Subjects, nearly evenly distributed among the three treatments, passed our manual screening. Now we report the exact instructions we used in T5. Those of the other two treatments were essentially the same, a part from obvious changes. Since the first three screens were basically the same as the first three screens in the first session, we start from Screen 4.

\textbf{Screen 4.} Benevolence is defined as giving a benefit to someone else at negligible cost to yourself. Please write a paragraph describing a time when acting benevolently led you in the right direction and resulted in a positive outcome for you. Alternatively, write a paragraph describing a time when acting malevolently led you in the wrong direction and resulted in a negative outcome for you. Of course, anything you write will be treated in the strictest confidence.

\textbf{Screen 5.} Now you are asked to make a decision. You are paired with another anonymous worker. You can earn a bonus depending on your and the other participant's decision. You and the other participant are both given 10c and you both have the same decision to make. You can either keep the 10c or give it to the other participant. In this case we will multiply that amount by 2 and the other participant will earn 20c. So, if you both keep the money, you will earn 10c each; if you both give all of your money, you will earn 20c each; if you keep all of your money and the other participant gives all of their money, you will earn 30c; if you give all of your money and the other participant keeps all of theirs, then you will earn nothing. The other person is REAL and will really make a decision.

\textbf{Screen 6.} What is your choice?

Here participants could select to either `Give' or `Keep'. No intermediate choices were allowed. After making their choice, subjects entered the demographic questionnaire. One of the questions was to describe the reason of their choice.


\begin{thebibliography}{1}
\bibitem{Tr} Trivers R (1971) The evolution of reciprocal altruism. \emph{Q Rev Biol} 46, 35-57.
\bibitem{Ax-Ha} Axelrod R, Hamilton WD (1981) The evolution of cooperation. \emph{Science} 211, 1390-1396.
\bibitem{MSK} Milinski M, Semmann D, Krambeck HJ (2002) Reputation helps solve the `tragedy of the commons'. \emph{Nature} 415, 424-426.
\bibitem{DH05} Doebeli M, Hauert C (2005) Models of cooperation based on the PrisonerÕs Dilemma and the Snowdrift game. \emph{Ecol Lett} 8, 748-766.
\bibitem{LK06} Lehmann L, Keller L (2006) The evolution of cooperation and altruism. A general framework and a classification of models. \emph{J Evol Biol} 19, 1365-1376.
\bibitem{No06} Nowak MA (2006) Five rules for the evolution of cooperation. \emph{Science} 314, 1560-1563.
\bibitem{C09} Crockett MJ (2009) The neurochemistry of fairness. \emph{Ann NY Acad Sci} 1167, 76-86.
\bibitem{Ra-No} Rand DG, Nowak MA (2013) Human cooperation. \emph{Trends Cogn Sci} 17, 413-425.
\bibitem{Ca} Capraro V (2013) A Model of Human Cooperation in Social Dilemmas. \emph{PLoS ONE} 8(8): e72427.
\bibitem{ZM} Zaki J, Mitchell JP (2013) Intuitive Prosociality. \emph{Curr Dir Psychol Sci} 22, 466-470.
\bibitem{BR} Boyd R, Richerson PJ (1992) Punishment allows the evolution of cooperation (or anything else) in sizable groups. \emph{Ethol Sociobiol} 13, 171-195.
\bibitem{FG1} Fehr E, G\"achter S (2000) Cooperation and punishment in public goods experiments. \emph{Am Econ Rev} 90, 980-994.
\bibitem{FG2} Fehr E, G\"achter S (2002) Altruistic punishment in humans. \emph{Nature} 415, 137-140.
\bibitem{GIR} G\"urerk \"O, Irlenbusch B, Rockenbach B (2006)The competitive advantage of sanctioning institutions. \emph{Science} 312, 108-111.
\bibitem{PB} Panchanathan K, Boyd R (2004) Indirect reciprocity can stabilize cooperation without the second-order free rider problem. \emph{Nature} 432, 499-502.
\bibitem{MSKM} Milinski M, Semmann D, Krambeck HJ, Marotzke J (2006) Stabilizing the Earth's climate is not a losing game: supporting evidence from public goods experiments. \emph{Proc Natl Acad Sci USA} 103, 3994-3998.
\bibitem{AHV} Andreoni J, Harbaugh W, Vesterlund L (2003) The carrot or the stick: rewards, punishments, and cooperation. \emph{Am Econ Rev} 93, 893-902.
\bibitem{RM} Rockenbach B, Milinski M (2006) The efficient interaction of indirect reciprocity and costly punishment. \emph{Nature} 444, 718-723.
\bibitem{SSW} Sefton M, Shupp R, Walker JM (2007) The effects of rewards and sanctions in provision of public goods. \emph{Econ Inq} 45, 671-690.
\bibitem{HS} Hilbe C, Sigmund K (2010) Incentives and opportunism: from the carrot to the stick. \emph{Proc R Soc B} 277 (1693), 2427-2433.
\bibitem{C} Cooper R, DeJong DV, Forsythe R, Ross TW (1996) Cooperation without Reputation: Experimental Evidence from Prisoner's Dilemma Games. \emph{Games Econ Behav} 12, 187-218.
\bibitem{GH} Goeree J, Holt C (2001) Ten Little Treasures of Game Theory and Ten Intuitive Contradictions. \emph{Am Econ Rev} 91, 1402-1422.
\bibitem{Ze} Zelmer J (2003) Linear public goods experiments: A meta-analysis. \emph{Exper Econ} 6, 299-310.
\bibitem{CJR} Capraro V, Jordan JJ, Rand DG (2014) Cooperation Increases with the Benefit-to-Cost Ratio in One-Shot Prisoner's Dilemma Experiments. Available at SSRN: http://ssrn.com/abstract=2429862.
\bibitem{RGN} Rand DG, Green JD, Nowak MA (2012) Spontaneous giving and calculated greed. \emph{Nature} 489, 427-430.
\bibitem{R} Rand DG, Peysakhovich A, Kraft-Todd GT, Newman GE, Wurzbacher O, Nowak MA, Greene JD (In press) Social Heuristics Shape Intuitive Cooperation. \emph{Nature Commun}.
\bibitem{PR} Peysakhovich A, Rand DG (2013) Habits of Virtue: Creating Norms of Cooperation and Defection in the Laboratory. Available at SSRN: http://ssrn.com/abstract=2294242.
\bibitem{RKT} Rand DG, Kraft-Todd GT (2013) Reflection Does Not Undermine Self-Interested Prosociality: Support for the Social Heuristics Hypothesis. Available at SSRN: http://ssrn.com/abstract=2297828.
\bibitem{DFR} Dreber A, Fudenberg D, Rand DG (2014) Who cooperates in repeated games: The role of altruism, inequity aversion, and demographics. \emph{J Econ Behav Organ} 98, 41-55.
\bibitem{HK} Harbaugh WT, Krause K (2000) Children's contributions in public goods experiments: the development of altruistic and free-riding behaviours. \emph{Econ Inq} 38, 95-109.
\bibitem{BEN} Blanco M, Engelmann D, Normann HT (2011) A within-subject analysis of other regarding preferences. \emph{Games Econ Behav} 72, 321-338.
\bibitem{BG06} Bra\~nas-Garza P (2006) Poverty in dictator games: Awakening solidarity. \emph{J Econ Behav Organ} 60, 306-320.
\bibitem{BG07} Bra\~nas-Garza P (2007) Promoting helping behavior with framing in dictator games. \emph{J Econ Psychol} 28(4):477-486.
\bibitem{CG08} Charness G, Gneezy U (2008) What's in a name?: anonymity and social distance in dictator and ultimatum games. \emph{J Econ Behav Organ} 68, 29-35.
\bibitem{E} Engel C (2011) Dictator games: A meta study. \emph{Exper Econ} 14, 583-610.
\bibitem{FP} Franzen A, Pointner S (2013) The external validity of giving in the dictator game. \emph{Exper Econ} 16, 155-169.
\bibitem{PCI} Paolacci G, Chandler J, Ipeirotis PG (2010) Running Experiments on Amazon Mechanical Turk. \emph{Judgm Decis Mak} 5, 411-419.
\bibitem{HRZ} Horton JJ, Rand DG, Zeckhauser RJ (2011) The online laboratory: conducting experiments in a real labor market. \emph{Exper Econ} 14, 399-425.
\bibitem{R12} Rand DG (2012) The promise of Mechanical Turk: How online labor markets can help theorists run behavioral experiments. \emph{J Theor Biol} 299, 172-179.
\bibitem{TSW}Thaler S, Simperl E, W\"olger S (2012) An experiment in comparing human-computation techniques. \emph{IEEE Internet Computing}, 16 (5):52-58. 
\bibitem{FS} Fehr E, Schmidt K (1999) A theory of fairness, competition, and cooperation. \emph{Q J Econ} 114 (3): 817-868. 
\bibitem{BO} Bolton GE, Ockenfels A (2000) ERC: A Theory of Equity, Reciprocity, and Competition. \emph{Am Econ Rev} 90 (1):166-193.
\bibitem{DF} Dawes CT, Fowler JH, Johnson T, McElreath R, Smirnov O (2007) Egalitarian motives in humans. Nature 446, 794-796.
\bibitem{Ch-Ra} Charness G, Rabin M (2002) Understanding social preferences with simple tests. \emph{Q J Econ} 117 (3): 817-869.
\bibitem{ES} Engelmann D, Strobel M (2004) Inequality aversion, efficiency, and maximin preferences in simple distribution experiments. \emph{Am Econ Rev} 94 (4):857-869.
\bibitem{CVPJ} Capraro V, Venanzi M, Polukarov M, Jennings NR (2013) Cooperative equilibria in iterated social dilemmas. \emph{In: Proc 6th of the International Symposium on Algorithmic Game Theory}, Lecture Notes in Computer Science 8146, pp. 146-158.
\bibitem{BC} Barcelo H, Capraro V (2014) Group size effect on cooperation in social dilemmas. Available at SSRN: http://ssrn.com/abstract=2425030
\bibitem{FF} Fehr E, Fischbacher U (2004) Social norms and human cooperation. \emph{Trends Cogn Sci} 8(4):185-190.
\bibitem{T13} Tomasello M, Vaish A (2013) Origins of Human Cooperation and Morality. \emph{Annu Rev Psychol} 64:231-255.
\bibitem{A}  Andreoni J (1989) Giving with Impure Altruism: Applications to Charity and Ricardian Equivalence. \emph{J Polit Econ} 97 (6): 1447-1458.
\bibitem{SW} Suri S, Watts DJ (2011) Cooperation and Contagion in Web-Based, Networked Public Goods Experiments. PLoS ONE 6(3):e16836.
\bibitem{RAC} Rand DG, Arbesman S, Christakis NA (2011) Dynamic social networks promote cooperation in experiments with humans. PNAS 108 (48) 19193-19198.
\end{thebibliography}
\end{document}